\documentclass[prd,aps,showpacs,nofootinbib,preprint,eqsecnum]{revtex4}
\usepackage{graphicx,color,amsmath,amsxtra}
\usepackage{amssymb}
\usepackage[english]{babel}
\usepackage{amsfonts}

\allowdisplaybreaks[4]
\begin{document}

\title{Cosmological perturbations in $k$-essence model}

\author{Kazuharu Bamba$^{1}$\footnote{E-mail address: bamba@kmi.nagoya-u.ac.jp}, 
Jiro Matsumoto$^{2}$\footnote{E-mail address: matumoto@th.phys.nagoya-u.ac.jp}, and 
Shin'ichi Nojiri$^{1, 2}$\footnote{E-mail address: nojiri@phys.nagoya-u.ac.jp}
}
\affiliation{
$^1$Kobayashi-Maskawa Institute for the Origin of Particles and the Universe, 
Nagoya University, Nagoya 464-8602, Japan\\
$^2$Department of Physics, Nagoya University, Nagoya 464-8602, Japan\\
}

\begin{abstract}

Subhorizon approximation is often used in cosmological perturbation theory. 
In this paper, however, it is shown that the subhorizon approximation is not always
a good approximation at least in case of $k$-essence model. 
We also show that the sound speed given by $k$-essence model 
exerts a huge influence on the time evolution of the matter density perturbation, 
and the future observations could clarify the differences between the $\Lambda$CDM model 
and $k$-essence model. 

\end{abstract}


\maketitle

\section{Introduction}

In order to explain the current accelerated expansion of the Universe, we need to modify 
the Einstein equation. 
There are two ways to modify the equation. 
One way is the modification of the right hand side of the Einstein equation by including unknown 
cosmological fluid, called ``dark energy''. A typical example of the dark energy is the $\Lambda$CDM model 
and $k$-essence model \cite{k-essence 1, k-essence 2, k-essence 3} 
is also one of the dark energy models. 
The $k$-essence model had been known as the model which causes the accelerated expansion of the Universe 
only by the kinetic terms of a scalar field. 
However, the $k$-essence model is currently recognized as a general single scalar field model which includes 
quintessence model \cite{quintessence 1, quintessence 2, quintessence 3, quintessence 4}, 
ghost condensate model \cite{ghost condensate}, tachyon model \cite{tachyon 1, tachyon 2, tachyon 3, tachyon 4} 
and so on. 
Another way to modify the Einstein equation is the modification of the left hand side 
of the Einstein equation, which is called ``modified gravity''. 
For example, scalar-tensor theory 
\cite{jordan, bd, Fujii_Maeda} and $F(R)$ gravity model 
\cite{Nojiri:2006ri, Sotiriou:2008rp, DeFelice:2010aj, Nojiri:2010wj} are known as 
the modified gravity models. 
Only from the data about the expansion rate 
of the Universe which is given by the observations of supernovae of type Ia, we cannot determine which model 
could be real one among such many models. 
One way to solve the problem is to consider the evolution of the matter density perturbation. 
The behavior of the evolution of the matter density perturbation is usually obtained by solving 
the equations of the linear perturbations, consisting of the metric perturbation and 
the perturbation of the energy momentum tensor. 
It is difficult, however, to solve the equations because scalar field models have more degrees of freedom 
than those in the $\Lambda$CDM model. 
Hence, the remaining way to evaluate the evolution of matter density perturbation has been 
using computers to solve the equations numerically under special conditions or 
to apply subhorizon approximation to the equations. 

In this paper, we will see that 
by linearizing equations without any other approximation, we find
the behavior of the evolution 
of the matter density perturbation in Sec. \ref{III}. 
We will also see that the matter density perturbation 
shows the oscillation whose frequency is proportional to the sound speed 
in $k$-essence model at the 
leading order of small scale and the effective growth factor of it is determined up to by the second derivatives 
respect to $X \equiv - \partial ^\mu \phi \partial _\mu \phi /2$ 
of the Lagrangian density $K(\phi, X)$ and $\dot \phi$ and their time derivatives. 
A clear observational difference between the $\Lambda$CDM and $k$-essence model could be also developed by the sound speed 
in $k$-essence model, which is also shown in Sec. \ref{III}. 
The last section is devoted to the conclusions and discussions.
We use units of $k_\mathrm{B} = c = \hbar = 1$ and denote the
gravitational constant $8 \pi G$ by
${\kappa}^2 \equiv 8\pi/{M_{\mathrm{Pl}}}^2$
with the Planck mass of $M_{\mathrm{Pl}} = G^{-1/2} = 1.2 \times 10^{19}$GeV.

\section{Cosmological perturbations}

\subsection{Equations of linear perturbations}

We start with the following action of $k$-essence model: 
\begin{align}
\label{10}
S=  \int d^4 x \sqrt{-g} \bigg \{ \frac{R}{2\kappa^2} - K(\phi, X) 
+ L_\mathrm{matter} \bigg \} ,\quad X \equiv - \frac{1}{2}\partial^\mu \phi \partial_\mu \phi\, .
\end{align}
Here $\phi$ is a scalar field and $L_\mathrm{matter}$ expresses the Lagrangian density of matters.  
Following the variational principle, by the functional differentiation with respect to metric $g^{\mu \nu}(x)$, 
we obtain the Einstein equations, 
\begin{align}
\label{20}
\frac{1}{\kappa ^2} G_{\mu \nu} = &- K(\phi , X) g_{\mu \nu} 
 - K_{,X} \partial _{\mu} \phi \partial _{\nu} \phi + T^{(m)}_{\mu \nu}\, ,
\end{align}
where $K_{,X}$ expresses the partial derivative of $K(\phi, X)$ with respect to $X$ 
and $T^{(m)}_{\mu \nu}$ is the energy-momentum tensor of the matters. 
On the other hand, the equation of motion for the scalar field $\phi$ is given by
\begin{align}
\label{30}
K_{,\phi} + \nabla _{\mu} (K_{,X} \partial ^{\mu} \phi) =0\, .
\end{align}
When the Universe is described by the 
Friedmann-Lemaitre-Robertson-Walker (FLRW) space-time with the flat spacial part, 
whose metric is given by $ds^2=-dt^2+ \sum_{i,j=1,2,3} \delta_{ij} a^2(t) dx^i dx^j$, 
Eqs.~(\ref{20}) and (\ref{30}) have the following forms:
\begin{align}
\frac{3H^2}{\kappa ^2} = & K- \dot \phi ^2 K_{,X} + \rho_\mathrm{matter}\, , \label{32} \\
 -\frac{2 \dot H}{\kappa ^2} = & - \dot \phi ^2 K_{,X} +(1+w) \rho _\mathrm{matter}\, , \label{35} \\
0 = & 3H \dot \phi K_{,X} + \ddot \phi (K_{,X} + \dot \phi ^2 K_{,XX}) -K_{, \phi} + \dot \phi ^2 K_{, X \phi} \, , 
\label{37}
\end{align}
where $w \equiv p_\mathrm{matter}/ \rho _\mathrm{matter} $ is the equation of state (EoS) 
parameter and $H(t)\equiv \dot a(t)/a(t)$. 
The energy momentum tensor is treated as that 
of perfect fluid, here. 
Then, the continuity equation for the matters is given by
\begin{equation}
\dot \rho _\mathrm{matter}+ 3(1+w) H \rho _\mathrm{matter}= 0\, . 
\label{39}
\end{equation}
In the following, 
we abbreviate the suffix ``matter'' for the 
quantities corresponding to matter, 
e.g., we write energy density $\rho _\mathrm{matter}$ by $\rho$, 
etc. 
We now consider the perturbed scalar field and the metric around the background, 
$\phi \rightarrow \phi + \delta \phi$, $X \rightarrow X+ \delta X$, 
and $g_{\mu \nu} \rightarrow g_{\mu \nu} + \delta g_{\mu \nu}$, to analyze 
the time evolution of the matter density perturbation. 
Then Eqs.~(\ref{20}) and (\ref{30}) give 
the following equations of linear perturbations, 
\begin{align}
\frac{1}{\kappa ^2} \delta R^{\ \nu}_{\mu} = & 
\frac{1}{2} \left ( \frac{1}{\kappa ^2} \delta R - K_{, \phi} \delta \phi 
 - K_{, X} \delta X \right ) \delta ^{\ \nu} _{\mu}
\nonumber \\
& - \frac{1}{2} \delta K_{,X} \partial _{\mu} \phi \partial ^{\nu} \phi 
 - \frac{1}{2} K_{,X} (\partial _{\mu} \delta \phi \partial ^{\nu} \phi 
+ \partial _{\mu} \phi \partial ^{\nu} \delta \phi)
+ \delta T^{(m)\, \nu}_{\mu}\, ,
\label{40}\\
0 = & K_{,\phi \phi} \delta \phi + K_{, \phi X}\delta X + \nabla _{\mu} (\delta K_{,X} \partial ^{\mu} \phi 
+ K_{,X} \partial _ {\alpha} \phi \delta g^{\alpha \mu}+ K_{,X} \partial ^{\mu} \delta \phi) \nonumber \\
& + \delta \Gamma ^{\mu} _{\mu \lambda} K_{,X} \partial ^{\lambda} \phi \, .
\label{50}
\end{align}
Here, we use the metric of FLRW space-time with the Newtonian gauge, 
\begin{align}
\label{60}
ds^2 = (-1+2 \Phi) dt^2 + \sum_{i,j=1,2,3} \delta _{ij} a^2(t) (1+2 \Psi) dx^i dx^j\, .
\end{align}
Then, Eq.~(\ref{40}) has the following forms: 
\begin{align}
-6H^2 \Phi -2 \frac{k^2}{a^2}\Psi -6H \dot \Psi =& \kappa ^2 \Big \{ 
(-K_{,X}+\dot \phi ^2 K_{,XX})(\dot \phi ^2 \Phi + \dot \phi \delta \dot \phi) \nonumber \\
&- (K_{, \phi} - \dot \phi ^2 K_{, X \phi} -2 \dot \phi K_{,X} \partial _0)\delta \phi 
+ 2 \Phi \dot \phi ^2 K_{,X} - \delta \rho \Big \}\, ,
\label{70} \\
2 \partial _i (H \Phi + \dot \Psi) =& 
\kappa ^2 \left \{ \dot \phi K_{,X} \partial _i \delta \phi + (\rho + p) \delta u_i \right \}\, ,
\label{80} \\
a^{-2} \partial _i \partial _j (\Phi - \Psi) =& 0, \quad (i \neq j)\, ,
\label{90} \\
\left( \frac{k^2}{a^2} + \frac{\partial _i \partial _i}{a^2} -2H \partial _0 -4 \dot H -6H^2  \right) \Phi
& - \left( \frac{k^2}{a^2} + \frac{\partial _i \partial _i}{a^2}  + 2 \partial _0 \partial _0 
+ 6H \partial _0 \right) \Psi  \nonumber \\
=& \kappa ^2 \left \{ -K_{,X} (\dot \phi ^2 \Phi + \dot \phi \delta \dot \phi)
 - K_{, \phi} \delta \phi + \delta p \right \}\, , \nonumber \\
&(\mbox{not summed with respect to $i$}),
\label{100} 
\end{align}
where Eqs.~(\ref{70})--(\ref{100}) express the ($0,0$), ($0,i$), ($i,j$) for $i \neq j$, and ($i,j$) for $i = j$ 
components of the Einstein equations, respectively.
In Eqs.~(\ref{70}) and (\ref{100}), $k$ expresses the wave number which appears from the derivative with respect 
to the spacial coordinates ($k^2 = - \partial _j \partial _j$) by the Fourier transformation. 

Since we have treated energy momentum tensor as perfect fluid 
$T_{\mu \nu} = pg_{\mu \nu} + (\rho +p)u_{\mu}u_{\nu}$, where $u_i =0$ and $u_0=-1$, 
the perturbation of the energy momentum tensor is given as follows, 
\begin{align}
\delta T^0_{\ 0} =& -\delta \rho\, , \label{110}\\
\delta T^0_{\ i} =& (\rho + p) \delta u_i\, , \label{120}\\
\delta T^i_{\ 0} =& - a^{-2} (\rho + p) \delta u_i\, , \label{130}\\
\delta T^i_{\ j} =& \delta^i_{\ j} \delta p\, , \label{140} 
\end{align}
where we have used the relation $\delta u_0 = \delta g _{00} /2$, 
which is obtained from the condition $g^{\mu\nu} u_\mu u_\nu = -1$. 
We now decompose $\delta u_i$ as $\partial _i \delta u + \delta u^V_i$, 
where $\delta u_i$ satisfies the transverse condition $\partial^i \delta u^V_i =0$. 
When we consider the scalar perturbation, we put $\delta u^V_i =0$ and keep only the scalar part $\delta u$. 
Then, $\delta u_i$ is expressed as $\delta u_i \equiv \partial _i \delta u$. 
On the other hand, perturbative equation of the scalar field (\ref{50}) has the following form: 
\begin{align}
(\dot \phi ^2 K_{,XX} + K_{,X}) \delta \ddot \phi 
=& - 3 \dot \phi K_{,X} \dot \Psi + (- \dot \phi K_{,X} \partial _0 - \dot \phi ^3 K_{,XX} \partial _0 
 -6H \dot \phi K_{,X} -2 \ddot \phi K_{,X} \nonumber \\
& -2 \dot \phi \dot K_{,X} -3H \dot \phi ^3 K_{,XX} -3 \dot \phi ^2 \ddot \phi K_{,XX}
  - \dot \phi ^3 \dot K_{,XX} + \dot \phi ^2 K_{, \phi X}) \Phi \nonumber \\
& - (3HK_{,X} + \dot K_{,X} +3H \dot \phi ^2 K_{,XX} +2 \dot \phi \ddot \phi K_{,XX} 
+ \dot \phi ^2 \dot K_{XX}) \delta \dot \phi \nonumber \\
& + \left ( -K_{,X} \frac{k^2}{a^2} -3H \dot \phi K_{,X \phi} - \dot \phi \dot K_{, X \phi} 
 - \ddot \phi K_{,X \phi} + K_{, \phi \phi} \right )\delta \phi\, . 
\label{165}
\end{align}
The perturbative equation of matter is given by linearizing 
the continuity equation $\nabla _{\mu} T^{\mu}_{\ \nu} =0$: 
\begin{align}
\delta \dot \rho + 3H(\delta \rho + \delta p) +a^{-2} \partial _i \{ (\rho +p) \delta u_i  \} 
+ 3 \dot \Psi (\rho + p) = 0\, , \label{150} \\
a^{-3} \partial _0 \{ a^3 (\rho + p) \delta u_i \} + \partial _i \delta p - (\rho + p) \partial _i \Phi 
=0\, . \label{160}
\end{align}
In the following, we consider the radiation dominant era, the matter dominant era, and the subsequent era and 
therefore we may assume $w= \mathrm{constant}$. 
In addition, we assume the sound speed $c_\mathrm{s} ^2 \equiv \delta p / \delta \rho$ behaves as constant. 

\subsection{Subhorizon approximation}

By using Eq.~(\ref{90}), we find we can choose $\Psi = \Phi$, which is assumed in the following. 
Subhorizon approximation is the approximation that $H^2$, $H \partial _0$, $\partial _0 ^2$ $\ll k^2 / a^2$. 
If we apply the subhorizon approximation in case of
$p=0$, $c_s ^2 = \delta p / \delta \rho = 0$ i.e. $\delta p = 0$, we obtain the following relation from
Eq.~(\ref{165}), 
\begin{align}
\vert \Psi \vert \gg \vert \delta \phi / \phi \vert\, . \label{260}
\end{align}
This relation implies we can neglect $\delta \phi$ in Eq.~(\ref{70}) 
and therefore we obtain the approximated 
equation from Eq.~(\ref{70}): 
\begin{align}
2 \frac{k^2}{\kappa ^2 a^2} \Psi \simeq \delta \rho\, . \label{250}
\end{align}
The above equation (\ref{250}) is identical with the equation of the Newton gravity.  
We can, then, obtain the evolution of matter perturbation by using Eqs.~(\ref{260}), 
(\ref{250}), (\ref{150}), and (\ref{160}).
Differentiating Eq.~(\ref{150}) with respect to $N \equiv \ln a$ 
and substituting Eq.~(\ref{160}), 
we obtain the following differential equation, 
\begin{equation}
\label{300}
\frac{d^2 \delta}{dN^2} + \left ( \frac{1}{2} - \frac{3}{2}w_\mathrm{eff} \right ) \frac{d \delta}{dN} 
 - \frac{3}{2} \Omega _\mathrm{m} \delta \simeq 0\, ,
\end{equation}
where $\delta \equiv \delta \rho / \rho$ and $\Omega _\mathrm{m}$ is the 
fractional density of matter, defined by $\kappa ^2 \rho  / (3H^2)$. 
$w_\mathrm{eff}$ is an effective EoS parameter defined by $w_\mathrm{eff} \equiv -1 - 2H'/3H$. 
This equation (\ref{300}) does not include the contribution from the scalar field $\phi$ explicitly, 
so that Eq.~(\ref{300}) is not changed from the equation given in the $\Lambda$CDM model. 
Thus, it might be considered that there is no contribution for the matter density perturbation explicitly from 
the scaler field $\phi$ when scaler field is only minimally coupled to the gravitation. 
This is, however, incorrect. Because the hypothesis we are able to use the subhorizon approximation is too 
strong to obtain a correct answer. The subhorizon approximation is decomposed into two approximations, 
that is, small scale approximation $k^2/a^2 \gg H^2$ and the assumption 
that the magnitude of the time variation of gravitational potential and the perturbation of the scalar field 
are almost that of the Hubble rate such as $\partial _0 \Psi \simeq H \Psi$, 
and the latter approximation can be only used in the case that 
the propagating speed of the perturbation mediated by the scalar field is negligible compared with the light speed. 
The behavior of the matter density perturbation in case that the contribution from the scalar field 
is not negligible in the subhorizon scale will be investigated in the next section. 

\section{Perturbative equations without approximations \label{III}}

In this section, we derive the differential equation of the matter perturbation $\delta$, 
without using any approximations except for using the simple Newtonian gauge metric perturbations 
and the perfect fluid approximation.

\subsection{Linear differential equations}

First, we rewrite the differential equations in the linearized forms.
After that, we will obtain a higher order differential equation
for the quantity $\delta \equiv \delta \rho / \rho$ in the next subsection.
In the following, we again assume $\Psi = \Phi$ from Eq.~(\ref{90}).
Then, Eq.~(\ref{70}) yields,
\begin{align}
& \dot \Psi - P_0 \delta \rho - P_1 \Psi - P_2 \delta X - P_3 \delta \phi = 0\, , \nonumber \\
& P_0 \equiv \frac{\kappa ^2}{6 H}\, , \quad 
P_1 \equiv -H - \frac{k^2}{3 a^2 H}\, , 
P_2 \equiv -\frac{\kappa ^2}{6H} \left( K_{,X} + \dot \phi ^2 K_{,XX} \right)\, , \nonumber \\
& P_3 \equiv \frac{\kappa ^2}{6H} \left( K_{, \phi} - \dot \phi ^2 K_{,X \phi} \right)\, . 
\label{1370} 
\end{align}
Eqs.~(\ref{165}) and (\ref{1370}) give, 
\begin{align}
\delta \dot X - E_0 & \delta \rho -E_1 \Psi -E_2 \delta X -E_3 \delta \phi=0\, , \nonumber \\
E_0 \equiv & - \frac{\kappa ^2}{2H} \dot \phi ^2 \frac{K_{,X}}{K_{,X} + \dot \phi ^2 K_{,XX}}\, , \nonumber \\
E_1 \equiv & \dot \phi (K_{,X} + \dot \phi ^2 K_{,XX})^{-1} \left \{ 3 \left ( H + 
 \frac{k^2}{3a^2 H} \right ) \dot \phi K_{,X} + \dot \phi ^2 K_{, \phi X} - K_{, \phi} \right \}\, , \nonumber \\
E_2 \equiv & \frac{\kappa ^2}{2H} \dot \phi ^2 K_{,X} - \frac{d}{dt} \ln 
 \left \vert \frac{a^3}{ \dot \phi} (K_{,X} + \dot \phi ^2 K_{,XX}) \right \vert \, ,  \nonumber \\
E_3 \equiv & - \dot \phi (K_{,X} + \dot \phi ^2 K_{,XX})^{-1} \nonumber \\
& \times \bigg \{ \frac{\kappa ^2}{2H}
\dot \phi K_{,X} (K_{, \phi} - \dot \phi ^2 K_{,X \phi} ) + \frac{d}{a^3 dt} (a^3 \dot \phi K_{, X \phi})
 + \frac{k^2}{a^2}K_{,X} - K_{, \phi \phi} \bigg \}\, .
\label{1380}
\end{align}
Eqs.~(\ref{80}), (\ref{150}), and (\ref{1370}) give, 
\begin{align}
\delta \dot \rho -& R_0 \delta \rho - R_1 \Psi - R_2 \delta X - R_3 \delta \phi = 0\, , \nonumber \\
R_0 & \equiv -3H(1+c_\mathrm{s}^2)+ \frac{\kappa ^2}{6H} \left \{ \frac{2k^2}{a^2 \kappa ^2} - 3(1+w) \rho \right \}\, , 
\nonumber \\
R_1 & \equiv - \frac{2k^4}{3a^4 H \kappa ^2} + 3(1+w) \rho \left ( H + \frac{k^2}{3a^2 H} \right )\, , \nonumber \\
R_2 & \equiv -\frac{\kappa ^2}{6H} \left \{  \frac{2k^2}{\kappa ^2 a^2} - 3(1 + w) \rho \right \}
\left ( K_{,X} + \dot \phi ^2 K_{,XX} \right )\, , \nonumber \\
R_3 & \equiv -\frac{k^2}{a^2} \dot \phi K_{,X} + \frac{\kappa ^2}{6H} \left \{  \frac{2k^2}{a^2 \kappa ^2}
- 3( 1 + w )\rho \right \} (K_{, \phi} - \dot \phi ^2 K_{,X \phi})\, , 
\label{1390}
\end{align}
where we have used $\delta p = c_\mathrm{s}^2 \delta \rho$. 
Therefore, from Eqs.~(\ref{1370})--(\ref{1390}), we find 
\begin{align}
\frac {d}{dt}
\left(
\begin{array}{c}
\delta \rho  \\
\Psi  \\
\delta X \\
\delta \phi 
\end{array}
\right) = 
\left(
\begin{array}{cccc}
R_0 & R_1 & R_2 & R_3 \\
P_0 & P_1 & P_2 & P_3 \\
E_0 & E_1 & E_2 & E_3 \\
0 & - \dot \phi & 1 / \dot \phi & 0
\end{array}
\right)
\left(
\begin{array}{c}
\delta \rho  \\
\Psi  \\
\delta X \\
\delta \phi 
\end{array}
\right)\, .
\label{1400}
\end{align}
We rewrite the above equations to the dimensionless forms for the later convenience: 
\begin{align}
\frac {d}{dN}
\left(
\begin{array}{c}
\delta \rho / \rho \\
\Psi  \\
\delta X / X \\
\delta \phi / \phi
\end{array}
\right) = 
\left(
\begin{array}{cccc}
A_{11} & A_{12} & A_{13} & A_{14} \\
A_{21} & A_{22} & A_{23} & A_{24} \\
A_{31} & A_{32} & A_{33} & A_{34} \\
A_{41} & A_{42} & A_{43} & A_{44}
\end{array}
\right)
\left(
\begin{array}{c}
\delta \rho / \rho \\
\Psi  \\
\delta X / X \\
\delta \phi / \phi
\end{array}
\right)\, .
\label{1410}
\end{align}
Here, $A_{ij}$ $\left(i,j=1,2,3,4\right)$ are given by
\begin{align*}
A_{11} \equiv & 3(w - c_\mathrm{s}^2)+ \frac{\kappa ^2}{6H^2} 
\left \{ \frac{2k^2}{a^2 \kappa ^2} - 3(1+w) \rho \right \}\, , \\
A_{12} \equiv & - \frac{2k^4}{3a^4 H^2 \kappa ^2 \rho} + 3(1+w) \left ( 1 + \frac{k^2}{3a^2 H^2} \right )\, , \\
A_{13} \equiv & - \frac{\kappa ^2}{12H^2} \left \{  \frac{2k^2}{\kappa ^2 a^2 \rho} - 3(1 + w) \right \}
 \dot \phi ^2 \left ( K_{,X} + \dot \phi ^2 K_{,XX} \right )\, , \\
A_{14} \equiv & \frac{k^2}{3a^2 H^2} \frac{1}{\rho} \phi \ddot \phi (K_{,X}+ \dot \phi ^2 K_{,XX})
 -(1+w) \frac{\kappa ^2}{2H^2} \phi (K_{, \phi}- \dot \phi ^2 K_{, X \phi})\, , \\
A_{21} \equiv & \frac{\kappa ^2 \rho}{6 H^2}\, , \quad 
A_{22} \equiv -1 - \frac{k^2}{3 a^2 H^2}\, , \\
A_{23} \equiv & -\frac{\kappa ^2}{12H^2} \dot \phi ^2 \left( K_{,X} + \dot \phi ^2 K_{,XX} \right), \quad
A_{24} \equiv \frac{\kappa ^2}{6H^2} \phi \left( K_{, \phi} - \dot \phi ^2 K_{,X \phi} \right)\, , \\ 
A_{31} \equiv & - \frac{\kappa ^2 \rho K_{,X}}{H^2( K_{,X} + \dot \phi ^2 K_{,XX})}\, , \\
A_{32} \equiv & \frac{2k^2}{a^2 H^2} \frac{K_{,X}}{K_{,X}+ \dot \phi ^2 K_{,XX}}- \frac{2 \ddot \phi}{H \dot \phi}\, , \\
A_{33} \equiv & -2 \frac{\ddot \phi}{H \dot \phi} + \frac{\kappa ^2}{2H^2} \dot \phi ^2 K_{,X} - \frac{d}{dN} \ln 
\left \vert \frac{a^3}{ \dot \phi} (K_{,X} + \dot \phi ^2 K_{,XX}) \right \vert\, , \\
A_{34} \equiv & - 2 \frac{\phi}{\dot \phi (K_{,X} + \dot \phi ^2 K_{,XX})} \bigg \{ \frac{\kappa ^2}{2H^2}
\dot \phi K_{,X} (K_{, \phi} - \dot \phi ^2 K_{,X \phi} ) \nonumber \\
& + \frac{1}{a^3} \frac{d}{dN} \left( a^3 \dot \phi K_{, X \phi} \right)
+ \frac{k^2}{a^2 H}K_{,X} - \frac{1}{H} K_{, \phi \phi} \bigg \}\, ,\\
A_{41} \equiv & 0, \quad
A_{42} \equiv -\frac{\dot \phi}{H \phi}\, , \quad
A_{43} \equiv \frac{\dot \phi}{2H \phi}\, , \quad
A_{44} \equiv - \frac{\dot \phi}{H \phi}\, .
\end{align*} 

\subsection{Evolution of the matter perturbation}

In the last subsection, we have derived a simultaneous differential equations for the perturbative quantities.
We will decompose the equations to give a single differential equation of the matter perturbation.
As a first step, we consider the case without the scalar field. 
Then, we obtain following equation from the first and the second lines of Eq.~(\ref{1410}),
\begin{align}
0 = &\frac{d^2 \delta}{dN^2} - \left ( A_{11}+A_{22}+ \frac{d}{dN} \ln \vert A_{12} \vert \right ) 
\frac{d \delta}{dN} \nonumber \\
& + \left ( A_{11} A_{22} - A_{12} A_{21} -\frac{dA_{11}}{dN} +A_{11} \frac{d}{dN} \ln \vert A_{12} 
\vert \right ) \delta \, .
\label{1420}
\end{align}
This is the differential equation for the matter density perturbation without approximation 
in the $\Lambda$CDM model \cite{lambda}. 
If we use the sound speed $c_\mathrm{s}$ considered subhorizon approximation, that is, 
not $\partial _0 \delta \simeq H \delta$ 
but $H^2 \delta \ll k^2 \delta / a^2$, Eq.~(\ref{1420}) is written as, 
\begin{align}
0 = &\frac{d^2 \delta}{dN^2} + \left \{ \frac{1}{2} -6w +3c_\mathrm{s}^2 -\frac{3}{2} w_\mathrm{eff} 
+ O \left ( \Big ( \frac{k^2}{a^2 H^2} \Big ) ^{-1} \right ) \right \} \frac{d \delta}{dN} \nonumber \\
& + \left \{ \frac{c_\mathrm{s}^2 k^2}{a^2 H^2} + 3(w-c_\mathrm{s}^2)(1+3w +3w_\mathrm{eff}) - \frac{3}{2}(1+3w)(1+w_\mathrm{eff}) 
+ O \left ( \Big ( \frac{k^2}{a^2 H^2} \Big ) ^{-1} \right ) \right \} \delta \, .
\label{1430}
\end{align}
Here, we have used $w_\mathrm{eff} \equiv -2 \dot H / 3H^2 -1$ and $(1+w)\kappa^2 \rho/ (3H^2) = 1+w_\mathrm{eff}$. 
Now we consider the case with the $k$-essence field $\phi$. From Eq.~(\ref{1410}), 
we obtain the fourth order differential equation,
\begin{align}
\frac{d^4 \delta}{dN^4} + M_3 \frac{d^3 \delta }{dN^3} 
+ M_2 \frac{d^2 \delta}{dN^2} + M_1 \frac{d \delta}{dN} + M_0 \delta = 0\, ,
\label{1440}
\end{align}
where the definition of the coefficients $M_3$, $M_2$, $M_1$, and $M_0$ are written in Appendix A. 
The expressions of the coefficients in the differential equation (\ref{1440}) are so complicated that 
it is difficult to analyze the differential equation without any approximations. 
If we use small scale approximation, $k^2/a^2 \gg H^2$, we can only consider the leading order terms in 
the expansion by the power series of $a^2 H^2 /k^2$ in the coefficients in Eq.~(\ref{1440}) 
as follows: 
\begin{align}
M_3 =& \bigg [ 7+3c_\mathrm{s}^2 -6w -\frac{2 \ddot \phi}{H \dot \phi} -3(1+w) \frac{\kappa ^2 \rho}{H^2} 
+ \frac{3 \kappa ^2 \dot \phi ^2 K_{,X}}{H^2} - \frac{6 K_{,X}}{4K_{,X}+ \dot \phi ^2 K_{,XX}}
\left ( 3 + \frac{\dddot \phi}{H \ddot \phi} \right ) \nonumber \\
&- \frac{\dot \phi (10 K_{,X}+\dot \phi ^2 K_{,XX})(\ddot \phi K_{,XX}+K_{, \phi X})}
{H K_{,X}(4K_{,X}+\dot \phi ^2 K_{,XX})} 
+ \frac{3 \dot \phi K_{,X}}{H \ddot \phi (K_{,X} + \dot \phi ^2 K_{, XX})(4K_{,X}+\dot \phi ^2 K_{,XX})} \nonumber \\
& \times \big \{ 3 (1+w)\kappa ^2 \rho K_{,X} -3 \kappa ^2 \dot \phi ^2 K_{,X}^2 +2K_{, \phi \phi}
 - 2 \ddot \phi K_{, \phi X}-6H \dot \phi K_{, \phi X}-2 \dot \phi ^2 K_{, \phi \phi X} \nonumber \\
& - 2 \dot \phi ^2 \ddot \phi K_{, \phi XX} \big \}  \bigg ]
 + O \left( \frac{a^2 H^2}{k^2} \right ) \, , \label{1460} \\
M_2 =& \frac{k^2}{a^2 H^2} \left ( c_\mathrm{s}^2 +\frac{K_{,X}}{K_{,X}+ \dot \phi ^2 K_{,XX}} \right )
 + O \left ( 1 \right ) \, ,
\label{1470} \\
M_1 =& \frac{k^2}{a^2 H^2} \Bigg [ c_\mathrm{s}^2 \left \{ -\frac{(1+w) \kappa ^2 \rho}{2H^2}
+ \frac{\kappa ^2 \dot \phi ^2 K_{,X}}{2H^2} - \frac{2 \ddot \phi}{H \dot \phi} \right \} \nonumber \\ 
& + \frac{K_{,X}}{K_{,X}+\dot \phi ^2 K_{,XX}} \left \{ 2 -6w 
 - \frac{(1+w) \kappa ^2 \rho}{2H^2} + \frac{\kappa ^2 \dot \phi ^2 K_{,X}}{2H^2} \right \} \nonumber \\
& + c_\mathrm{s}^2  \frac{-2 K_{,X} + \dot \phi ^2 K_{,XX} -6 \frac{\dddot \phi }{H \ddot \phi}K_{,X}}
{4 K_{,X} + \dot \phi ^2 K_{,XX}}
 - c_\mathrm{s}^2 \frac{\dot \phi (10 K_{,X}+ \dot \phi ^2 K_{,XX})(\ddot \phi K_{,XX} + K_{, \phi X})}
{H K_{,X}(4 K_{,X} + \dot \phi ^2 K_{,XX})} \nonumber \\
& + c_\mathrm{s}^2 \frac{3 K_{,X}}
{H \ddot \phi (K_{,X}+\dot \phi ^2 K_{,XX})(4 K_{,X}+ \dot \phi ^2 K_{,XX})}
 \left\{ -3H \dot \phi ^2 (\ddot \phi K_{,XX} +2K_{, \phi X}) \right. \nonumber \\
& + 3(1+w) \kappa ^2 \rho \dot \phi K_{,X} -3 \kappa ^2 \dot \phi ^3 K_{,X}^2 
-2 \dot \phi \ddot \phi K_{,\phi X} +2 \dot \phi K_{, \phi \phi} \nonumber \\
& \left. -2 \dot \phi ^3 \ddot \phi K_{, \phi XX} -2 \dot \phi ^3 K_{, \phi \phi X} \right\} \Bigg ]
+ O \left ( 1 \right )\, , \label{1480} \\
M_0 =& \frac{k^4}{a^4 H^4} \left ( \frac{c_\mathrm{s}^2 K_{,X}}{K_{,X}+ \dot \phi ^2 K_{,XX}} \right )
 + O \left ( \frac{k^2}{a^2 H^2} \right )\, . 
\label{1490} 
\end{align}
Here, we have deleted $\dot H$, $\dot \rho$, and the higher derivatives of them 
by using Eqs.~(\ref{35}) and (\ref{39}). 
Furthermore, we have used the field equation (\ref{37}) to delete $K_{, \phi}$ and also used derivatives of Eq. (\ref{37}) 
to delete $K_{, XXX}$, $K_{, XXXX}$, and $K_{, XXXXX}$. 
Thus, these coefficients (\ref{1460})--(\ref{1490}) can be expressed 
in different forms by using the background equations 
(\ref{32})--(\ref{39}) and their derivatives.  

If we use the WKB approximation under the condition, $k^2/a^2 \gg H^2$, 
we can obtain the solution of Eq.~(\ref{1440}) as follows, 
\begin{align}
\delta (N) = \sum_{i=1}^4 C_i(N) \exp \left [\int ^N dN' \frac{\lambda_i k}{a H} \right]\, . 
\label{1500}
\end{align}
Here, $C_i(N)$'s are functions determined by the leading terms in Eqs.~(\ref{1460})--(\ref{1480}) as the terms 
proportional to $k^3/(aH)^3$ in Eq. (\ref{1440}), and 
$\lambda_i$'s are determined by the terms proportional to $(k/aH)^4$ as 
$\lambda_1=- i c_\mathrm{s}$, $\lambda_2 = i c_\mathrm{s}$, $\lambda_3 = - i c_\phi$, 
and $\lambda_4 = i c_\phi$, respectively. 
The sound speed in $k$-essence model $c_\phi ^2 \equiv (p_{\phi})_{,X} / (\rho_{\phi})_{,X}$ \cite{k-essence 2} 
is given by $K_{,X}/(K_{,X}+ \dot \phi ^2 K_{,XX})$, where $\rho_\phi$ and $p_\phi$ are the energy density and pressure 
of the $k$-essence scalar field, respectively. 
If $c_\mathrm{s}$ is a real number and $c_\phi (\phi, X)$ is a real function, then  the solution (\ref{1500}) oscillates rapidly 
in the small scale $k/a \gg H$ since $\lambda_i$'s are pure imaginary. 
We will investigate this oscillation later. 
If we require the solution (\ref{1500}) to be a real function, we should replace it with the following form: 
\begin{align}
\delta (N) = &C_1(N)  \cos \left [\int ^N dN' \frac{c_s k}{a H} \right] 
+ C_2(N) \sin \left [\int ^N dN' \frac{c_s k}{a H} \right] \nonumber \\
&+ C_3(N) \cos \left [\int ^N dN' \frac{c_\phi k}{a H} \right]
+ C_4(N) \sin \left [\int ^N dN' \frac{c_\phi k}{a H} \right]\, , 
\label{1510}
\end{align}
where $C_1(N)$, $C_2(N)$, $C_3(N)$, and $C_4(N)$ are real functions of $N$. 
$C_1(N)$ and $C_2(N)$ are expressed as follows by to solve the third order equation of $k/(aH)$ of Eq. (\ref{1440}), 
\begin{equation}
\frac{d}{dN} \ln \vert C_1(N) \vert = \frac{d}{dN} \ln \vert C_2(N) \vert = \frac{1}{2}(-1+6w-3c_\mathrm{s} ^2). 
\label{1512}
\end{equation}
The above expression gives $d \ln \vert C_1(N) \vert /dN =
 d \ln \vert C_2 (N) \vert /dN \leq  0$, therefore, 
these are not growing modes of the matter density perturbation. 
The behavior of these solutions $C_1(N) \cos \left [\int ^N dN' c_s k/(a H) \right]$ and 
$C_2(N) \sin \left [\int ^N dN' c_s k/(a H) \right]$ are completely the same behavior of the solutions
of Eq. (\ref{1430}) when $c_s \neq 0$. 
In the following, we will focus on the matter dominant era and the subsequent era, then 
the equation of state for matter and the sound speed of matter can be treated as $w=0$ 
and $c_\mathrm{s}=0$, respectively. 
The term proportional to $k^4/(a^4 H^4)$ in Eq. (\ref{1490}), then, vanishes, and 
we should evaluate more lower order terms. 
When $c_\mathrm{s}^2 = w = 0$, $M_0$ is expressed 
as 
\begin{align}
M_0' =& -c_\phi ^2 \frac{k^2}{a^2 H^2}\frac{3}{2}\Omega_\mathrm{m} \delta
 + O \left (  1 \right ) \, .
\label{1514}
\end{align}
For $M_0'$ in (\ref{1514}), 
$C_1(N) \cos \left [\int ^N dN' c_s k/(a H) \right]$ and 
$C_2(N) \sin \left [\int ^N dN' c_s k/(a H) \right]$ are not 
the solutions of Eq.~(\ref{1440}). 
In the following subsections, we investigate the solutions and estimate 
the time evolution of $C_3(N)$ and $C_4(N)$ in case of quintessence model and general $k$-essence model. 

\subsection{Quintessence model}

When we restrict the $k$-essence action to be the action of quintessence model ($K(\phi, X)= -X + V(\phi)$), 
the forms of functions $C_3(N)$ and $C_4(N)$ are given by  
\begin{equation}
\frac{d}{dN} \ln \vert C_3(N) \vert = \frac{d}{dN} \ln \vert C_4(N) \vert 
= -\frac{3}{2} + \frac{\ddot \phi}{H \dot \phi } - \frac{\dddot \phi}{2H \ddot \phi} 
 - \frac{\dot \phi}{2H \ddot \phi} \left (  3 \dot H + V_{,\phi \phi} \right )\, . 
\label{1520}
\end{equation}
We can simplify this expression (\ref{1520}) by using the equation 
of motion for the quintessence model, $\ddot \phi + 3H \dot \phi +V_{,\phi}=0$ as follows: 
\begin{equation}
\frac{d}{dN} \ln \vert C_3(N) \vert = \frac{d}{dN} \ln \vert C_4(N) \vert 
= \frac{\ddot \phi}{H \dot \phi}\, . 
\label{1530}
\end{equation}
Then we find $C_3(N)$, $C_4(N) \propto \dot \phi(N)$. 
The effective evolution of the growth factor $d \ln \delta (N) / dN$ 
is determined by the evolution of $f_\mathrm{eff} \equiv d \ln \vert C_3(N) \vert / dN$ = 
$d \ln \vert C_4(N) \vert / dN$ = $\ddot \phi / ( H \dot \phi )$ 
because the exponential factor of the solution 
(\ref{1510}) is oscillating. 
If we consider a quintessence model with an exponential potential, 
which is $V(\phi)=B^2 {\mathrm{e}}^{- \sqrt{3/2}\kappa \phi}$, 
we know the exact solutions \cite{C.Rubano P.Scudellaro}, 
\begin{align}
a^3(t)= & (u_1 t + u_2)\left (\frac{1}{4} u_1 \kappa^2 B^2 t^3 
+ \frac{3}{4} u_2 \kappa^2 B^2 t^2 +v_1 t + v_2 \right)\, , 
\label{1540} \\
\phi (t) = & - \frac{\sqrt{2}}{\sqrt{3} \kappa} \ln \frac{u_1 t + u_2}
{\frac{1}{4} u_1 \kappa^2 B^2 t^3 
+ \frac{3}{4} u_2 \kappa^2 B^2 t^2 +v_1 t + v_2}\, , \label{1550}
\end{align}
where $u_1$, $u_2$, $v_1$, and $v_2$ are arbitrary constants of integration. 
We now impose $u_2 = v_2 =0$ to satisfy the conditions $a(0)=0$ and $\Omega_\phi (0)=0$ 
because we do not consider the inflationary regime here. 
Then the effective growth factor $f_\mathrm{eff}=\ddot \phi /(H \dot \phi)$ is given by 
\begin{align}
f_\mathrm{eff}(t) = -\frac{3}{4} + \frac{9}{2} \frac{v_1}{u_1 \kappa ^2 B^2  t^2 + 2 v_1}\, . 
\label{1560}
\end{align}
Here, the constants $u_1$ and $v_1$ are restricted to satisfy $\rho_{0} = 4 u_1 v_1 /3 \kappa ^2$ 
by Eq.~(\ref{32}) when the present scale factor $a_0$ is chosen to be unity, 
so that $u_1 v_1 > 0$ because $\rho_{0} > 0$. 
Therefore, $f_\mathrm{eff}(t)$ is a monotonically decreasing function 
and the value of $f_\mathrm{eff}$ is restricted to be $-3/4 \leq f_\mathrm{eff} \leq 3/2$ 
because $\lim _{t \rightarrow 0} f_\mathrm{eff} = 3/2$ and 
$\lim _{t \rightarrow \infty} f_\mathrm{eff} = -3/4$. 
The current value of $f_\mathrm{eff}$ is estimated by using the parameter $t_s= 15.8 \times 10^9$ years 
defined by $t_s ^2 \equiv 4 v_1 / \kappa ^2 B^2 u_1$. The value of $t_s$ follows from the conditions 
$\Omega _{m0} = 0.3$ and $H_0=70\, \mathrm{km\, s}^{-1}\mathrm{Mpc}^{-1}$ 
\cite{C.Rubano P.Scudellaro}. 
With the present time $t_0 = 13 \times 10^9$ years, we find $f_\mathrm{eff}(t_0) \thickapprox 0.21$, which does not 
contradict with the observational results \cite{Guzzo et al.}. 

On the other hand, the remaining two solutions in four linearly independent solutions 
are determined as follows. 
In quintessence model, the coefficient $M_0'$ is expressed as 
\begin{align}
M_0' =& -\frac{k^2}{a^2 H^2}\frac{3}{2}\Omega_\mathrm{m} \delta
 + O \left (  1 \right ) \, .
\label{q1}
\end{align}
By taking account of 
the second order terms of $k/(aH)$ of coefficients of Eq. (\ref{1440}), 
we find the following equation,  
\begin{align}
\frac{d^2 \delta}{dN^2} + \left( \frac{1}{2} - \frac{3}{2}w_\mathrm{eff} \right ) \frac{d \delta}{dN} 
-\frac{3}{2} \Omega_\mathrm{m} \delta = 0,  
\label{q2}
\end{align}
which will yield quasi-static solutions. This equation (\ref{q2}) is same as Eq.~(\ref{300}). 
In case of exponential potential quintessence model we have considered, 
solutions of Eq.~(\ref{q2}) are sub-leading ones because the growth rate of them are less than $3/2$ in matter dominant era.  

\subsection{General $k$-essence model}

In case of general $k$-essence model, Eq.~(\ref{q2}) is written as:
\begin{align}
c_\phi ^2 \left \{ \frac{d^2 \delta}{dN^2} + \left( \frac{1}{2} - \frac{3}{2}w_\mathrm{eff} \right ) \frac{d \delta}{dN} 
-\frac{3}{2} \Omega_\mathrm{m} \delta \right \}= 0.  
\label{k1}
\end{align}
This equation (\ref{k1}) equals to Eq.~(\ref{300}) when $c_\phi \neq 0$ i.e. $K_{,X} \neq 0$. 

On the other hand, the coefficient of the oscillating solution $C(N)$ is determined as 
\begin{align}
\frac{d}{dN} & \ln \vert C_3(N) \vert = \frac{d}{dN} \ln \vert C_4(N) \vert \nonumber \\
=&\frac{\ddot \phi}{H \dot \phi} 
 - \frac{5}{4} \frac{\dot \phi K_{, \phi X} + \dot \phi \ddot \phi K_{, XX}}{H K_{,X}}
+ \frac{\dot \phi (10 K_{, X}+ \dot \phi ^2 K_{, XX})(\ddot \phi K_{, XX}+K_{, \phi X})}
{2H K_{, X}(4 K_{, X}+ \dot \phi ^2 K_{, XX})} \nonumber \\
&+ \left \{ \frac{5}{4 H (K_{,X}+ \dot \phi ^2 K_{,XX})} 
 - \frac{3  K_{, X}}{H (K_{,X}+ \dot \phi ^2 K_{, XX})(4 K_{, X} + \dot \phi ^2 K_{, XX})} \right \}
\nonumber \\
& \times \left ( \dot \phi K_{, \phi X} +3 \dot \phi \ddot \phi K_{, XX} 
+ \dot \phi ^3  K_{, \phi XX} + \dot \phi ^3 \ddot \phi K_{, XXX} \right )\, . 
\label{2000}
\end{align}
Here, the derivative of the field equation (\ref{37}) is also used to simplify the equation. 
We will discuss the possibility of ``reconstruction'' of the effective growth factor 
$f_\mathrm{eff}= d \ln \vert C_3 (N) \vert / dN = d \ln \vert C_4 (N) \vert / dN$ in the following.  
Before considering the reconstruction of the effective growth factor $f_\mathrm{eff}$, we 
review on the reconstruction of background geometry. 
The reconstruction, we call here, is the method for constructing the action which yields some given behavior of 
quantities, e.g., $a(t)$, $H(t)$, or $f(N) \equiv d \ln \delta (N)/ dN$. 
We can always express $\phi (t)$ as $\phi (t) = g(t)$ by using some bijection $g(t)$ because
the redefinition of $\phi (t)$ can be absorbed into the redefinition of the function $K(\phi ,X)$. 
If we choose $g(t)$ as $\phi = t$, the action of $k$-essence model is reconstructed for arbitrary 
growth of the background space-time (for an arbitrary $a(t)$ or $H(t)$) as follows \cite{J.M. & S.N. }: 
\begin{align}
K(\phi, X) =& K^{(n)}(\phi ) \left ( X - \frac{1}{2} \right )^n\, , \nonumber \\
K^{(0)} (\phi) \equiv& \left. \left ( w \rho + \frac{3 H^2}{\kappa ^2} 
+ \frac{2 \dot H}{\kappa ^2} \right ) \right|_{t=\phi}\, , \nonumber \\
K^{(1)} (\phi) \equiv& \left. \left \{ \frac{2 \dot H}{\kappa ^2} +(1+w) \rho \right \} 
\right|_{t=\phi}\, . 
\label{2010}
\end{align}
The terms which are proportional to the square or higher power of $(X-1/2)$ do not contribute 
to the background evolution of space-time. 
Therefore, we can choose their coefficients $K^{(n)}(\phi)$ with $(n>1)$ as arbitrary functions of $\phi$. 

We now consider that the possibility of reconstruction of the effective growth factor 
by using the arbitrariness of $K^{(2)}(\phi)$. 
First, we should note that Eq.~(\ref{2000}) is rewritten as follows: 
\begin{align}
\frac{d}{dN} \ln \vert C(N) \vert =& \frac{d}{dN} \ln \vert \dot \phi 
\vert - \frac{3}{4} \frac{d}{dN} \ln \vert K_{, X} \vert
\nonumber \\ 
&+ \frac{1}{4} \frac{d}{dN} \ln \left\vert K_{,X} + \dot \phi ^2 K_{, XX} \right\vert 
+ \frac{d}{dN} \ln \vert 4K_{, X} + \dot \phi ^2 K_{, XX} \vert\, , 
\label{2020}
\end{align}
where we have used $C(N) \propto C_3 (N) \propto C_4 (N)$ as shown in (\ref{2000}). From 
the above equation (\ref{2020}), 
we immediately obtain the following relation:   
\begin{equation}
C(N) ^4 K_{, X}^3 = \mathrm{const.} \times \dot \phi ^4 (K_{, X} + \dot \phi ^2 K_{, XX})(4 K_{, X} 
+ \dot \phi ^2 K_{, XX})^4\, .
\label{2030}
\end{equation}
The reconstruction of the effective growth factor $f_\mathrm{eff}$ is completed if $f_\mathrm{eff}$ is integrable 
and Eq.~(\ref{2030}) is algebraically solvable for $K_{,XX}$ because $C(N)$ is expressed by using 
the integration of $f_\mathrm{eff}$. 
However, we cannot complete the reconstruction because Eq.~(\ref{2030}) is a quintic 
algebraic equation with respect to $K_{, XX}$ and hence it is not generally solvable.  
It should be noted that it is possible to execute the reconstruction approximately. 
By redefining $\alpha (N) \equiv \dot \phi^2 K_{, XX}/K_{, X}$ and 
$y(N) \equiv C^4(N) /(\mathrm{const.} \times \dot \phi ^4 K_{, X}^2) $, we can rewrite Eq.~(\ref{2030}) as 
\begin{equation}
y(N)=(1+ \alpha (N))(4+ \alpha (N))^4\, . 
\label{2040}
\end{equation}
Thus, we find that there are one or more real solutions for $\alpha$. 
To be concrete, in case $y<-2^8 3^5/5^5$ or $0<y$ there is one real solution 
for Eq.~(\ref{2040}), 
in case $y=-2^8 3^5/5^5, 0$ there are two real solutions, and 
in case $-2^8 3^5/5^5 < y < 0$ there are three real solutions. 
So that it is possible to define $\alpha (N)$ for the arbitrary behavior of $y(N)$. 
It should be, however, noted that the function $\alpha (N)$ must be a multi-valued function or a discontinuous function
if we consider the continuous function $y(N)$ which crosses over the region $[ -2^8 3^5/5^5, 0 ]$. 
\subsection{Oscillation}

Since we find the ratio between $C_3(N)$ and $C_4(N)$ is a constant factor from Eq.~(\ref{2000}), 
we can rewrite the oscillating solutions in the following form: 
\begin{align}
\delta _\mathrm{o}(N) = 
C(N) \sin \left [ \int ^N dN' \frac{c_\phi k}{a H} + \omega \right ]\, , 
\label{2050}
\end{align}
where $\omega$ is an arbitrary real constant. 
The frequency of the solution (\ref{2050}) is given by $ck/(a_0 H_0) \thickapprox 300$ when we use the values, 
$H_0 = h \times (9.777752 \, \mathrm{Gyr})^{-1}$ and $k= 0.1 h\, \mathrm{Mpc}^{-1}$. 
This corresponds to the frequency around the present time in quintessence model because $c_\phi = c$ is satisfied in 
quintessence model. 
We now examine how often the matter density perturbation becomes $0$. 
We set $300 \Delta z \thickapprox \pi$ and obtain $\Delta z \thickapprox \pi / 300$ 
because function $\sin\theta$ vanishes in the period of $\theta$ by $\pi$. 
There are points that the matter density perturbation vanishes, for the period $0.01$ of the redshift $z$. 
Generally, the frequency of the solution (\ref{2050}) is given by $ck(1+z')/(a_0 H(z=z'))$ when we consider it 
around $z=z'$. 
The difference of $N$, $\Delta N (z') \equiv N(z' - \Delta z) - N(z')$, is 
expressed as $\ln(1+z')/(1+z' - \Delta z)$ due to the definition of $N$. 
Hence, $\Delta N \thickapprox \Delta z /(1+z')$ is justified when $\vert 1+z' \vert \gg \vert \Delta z \vert$. 
The frequency corresponding to $z$, then, is expressed as $ck/(a_0 H(z=z'))$. 
The difference of the frequency corresponding to $z$ between around $z=0$ and around $z=z'$ is only $H(z)$. 
If we use a scale factor of the $\Lambda$CDM model, 
$a(t) = (\rho_{m0}/\rho _{\Lambda})^{1/3} \sinh ^{2/3} [\sqrt{3 \kappa ^2 \rho_{\Lambda}/4} \: t]$, 
we can evaluate the concrete value of the frequency because the ratio $H(z=z')/H(z=0)$ is determined. 
In case of $z=1$ and $k=0.01 h\, \mathrm{Mpc}^{-1}$, the frequency is approximately one seventeenth times of 
that in case of $z=0$ and $k=0.1 h \mathrm{Mpc}^{-1}$ because the ratio $H(z=1)/H(z=0)$ is approximately determined $1.7$. 
Here, we have used $\Omega _\Lambda = 0.74$, $\Omega _{m0} = 0.26$, and $h=0.72$. 
Thus, the distance of the points that the matter density perturbation vanishes becomes bigger than that in case of 
$z=0$ and $k= 0.1 h \, \mathrm{Mpc}^{-1}$, and the points appears in the interval of $0.18$ in the redshift. 

We now consider the variation of the wave number $k$ with a fixed redshift. 
The frequency corresponding to $k$, then, is expressed by $\int ^N dN' c_\phi / (aH)$. 
It can be replaced with $\int^t dt c_\phi / a(t)$ if we use $dN = H(t) dt$. 
Here, we assume the bottom point of the domain of the integration as a time of the radiation-matter equality 
$t_\mathrm{eq}$. We also investigate it in case of $c_\phi = c$ and 
$a(t)= (\rho_{m0}/\rho _{\Lambda})^{1/3} \sinh ^{2/3} [\sqrt{3 \kappa ^2 \rho_{\Lambda}/4} \: t]$ in order 
to evaluate the value of the frequency corresponding to $k$. 
The present value of the frequency is given by 
$c \int^{t_0} _{t_\mathrm{eq}} dt/a(t) \thickapprox 1.45 \times 10^4\, \mathrm{Mpc}$, where we have used 
$t_0 = 13.7 \times 10^9$ years, $t_\mathrm{eq}= 4000$ years, and the sidereal year equals $31558149.8\, \mathrm{s}$. 
So that, the amount of the variation of $k$ in order to make a phase shift with $\pi$ is 
$\Delta k \thickapprox 2.17 \times 10^{-4}\, \mathrm{Mpc}^{-1}$. 
It is so small that the observational verification should be very difficult. 
On the other hand, $\Delta k$ in $z=1$, which is given by setting the upper end of the domain of the integration 
to be $t= 6.0 \times 10^9$ years, is approximately $1.3$ times bigger than that in $z=0$, but 
it is also very small.


\section{Conclusions and Discussions}

We have estimated the time evolution of the matter density perturbation without using subhorizon approximation 
but using only small scale approximation $k/(aH) \gg 1$ in this paper. 
As a result, the characteristic behaviors of the matter density perturbation, the oscillation caused by the 
sound speed in $k$-essence model, and the effective growth factor depends on $K_{,X}$, $K_{,XX}$, $\dot \phi$, 
and their derivatives, have been discovered. 
From a viewpoint of verification of dark energy, the effective growth factor is not very important 
because even the simple example of quintessence model does not induce a great difference from the $\Lambda$CDM model, 
and the possibility of approximate reconstruction of the effective growth factor has been shown. 
However, the oscillation caused by the sound speed in $k$-essence model is a peculiar behavior which is 
not explained by the $\Lambda$CDM model, so that we can find a clear difference in it. 
As seen in the last section, the oscillation could be, moreover, observable if we fix the wavenumber $k$ and 
move the redshift $z$ around $k=0.01h \mathrm{Mpc}^{-1}$ and $z=1$. 
On the other hand, second order differentiated equation of the matter density perturbation which will give 
quasi-static solutions has been found, too. This differential equation is same as that, for which the subhorizon approximation 
have been applied when $c_\phi \neq 0$. 
Therefore, there is a possibility that the quasi-static solutions have a growing mode but the oscillating solutions do not 
have a growing mode, and thus the oscillation could not be able to be observed. 
Several conditions, however, will be imposed on the action to realize such a situation.
Consequently, if such an oscillation is not observed, $k$-essence model would be severely constrained. 
On the other hand, if such an oscillation is observed, the existence of the scalar field dark energy such as $k$-essence 
model or the dark matter given by a scalar field may be verified. 

\section*{Acknowledgments}

This work was supported in part by the Global COE Program of Nagoya University 
``Quest for Fundamental Principles in the Universe (QFPU)'' from JSPS and MEXT of Japan 
and the JSPS Grant-in-Aid for Scientific Research (S) \# 22224003  and (C) \# 23540296 (S.N.).

\appendix
\section{Derivation of the forth order differential equation}
The fourth order differential equation (\ref{1440}), 
\begin{align}
\frac{d^4 \delta}{dN^4} + M_3 \frac{d^3 \delta }{dN^3} 
+ M_2 \frac{d^2 \delta}{dN^2} + M_1 \frac{d \delta}{dN} + M_0 \delta = 0\, ,
\label{A}
\end{align}
which is given by the Gaussian elimination like procedure: 
\begin{enumerate}
\item Differentiating the first line of Eq.~(\ref{1410}) with respect to $N$. 
\item Eliminating the terms proportional to $d \Psi / dN$, $d (\delta X/ X) / dN$, 
and $d (\delta \phi / \phi)/ dN$ 
from the equation derived in the first step by using the second, third, and fourth lines of Eq.~(\ref{1410}). 
\item Differentiating the equation given in the second step with respect to $N$. 
\item Eliminating the terms proportional to $d \Psi / dN$, $d (\delta X/ X) / dN$, and $d (\delta \phi / \phi)/ dN$ 
from the equation derived in the third step by using the second, third, and fourth lines of Eq.~(\ref{1410}). 
\item Differentiating the equation given in the fourth step with respect to $N$. 
\item Eliminating the terms proportional to $d \Psi / dN$, $d (\delta X/ X) / dN$, and $d (\delta \phi / \phi)/ dN$ 
from the equation derived in the fifth step by using the second, third, and fourth lines of Eq.~(\ref{1410}). 
Then we obtain the fourth order differential equation for $\delta$ including the extra quantities 
$\Psi$, $\delta X / X$, and $\delta \phi / \phi$.
\item By using the equations obtained in the second and fourth steps and the first line of Eq.~(\ref{1410}), 
we can delete the terms proportional to $\Psi$, $\delta X / X$, and $\delta \phi / \phi$ in the equation 
obtained in the sixth step, and then we find Eq.~(\ref{A}). 
\end{enumerate}
Eq.~(\ref{A}) is the differential equation for the matter density perturbation in $k$-essence model without 
any approximation. 
The coefficients in (\ref{A}) are defined by 
\begin{align}
M_3 \equiv &-A_{11} 
+ \sum_{\alpha , \beta , \gamma =2\, \&\, \alpha \neq \beta \neq \gamma}^4 D(\alpha , \beta , \gamma)
A_{1 \beta} \bigg ( A_{1 \beta} \frac{dA_{1 \gamma}}{dN} \nonumber \\ 
& + A_{1 \beta} \sum ^4 _{i=2} A_{1i}A_{i \gamma} 
 - A_{1 \gamma} \frac{dA_{1 \beta}}{dN} - A_{1 \gamma} \sum ^4 _{i=2} A_{1i}A_{i \beta} \bigg )\, , \\ 
M_2 \equiv &- 3 \frac{dA_{11}}{dN} - \sum ^4 _{\alpha = 2} A_{1 \alpha}A_{\alpha 1} 
 - \sum_{\alpha , \beta , \gamma =2 \& \alpha \neq \beta \neq \gamma}^4 D(\alpha , \beta , \gamma)
A_{1 \beta} \bigg ( A_{1 \beta} \frac{d^2 A_{1 \gamma}}{dN^2} \nonumber \\
& + A_{11} A_{1 \beta} \frac{dA_{1 \gamma}}{dN} 
+ A_{11}A_{1 \beta} \sum ^4 _{d=2} A_{1d}A_{d \gamma} 
+ 2A_{1 \beta} \sum ^4 _{d=2} \frac{dA_{1 d}}{dN}A_{d \gamma}  \nonumber \\
& +A_{1 \beta} \sum ^4 _{d=2} A_{1d} 
\frac{dA_{d \gamma}}{dN} + A_{1 \beta} \sum ^4 _{d,e=2} A_{1e}A_{ed}A_{d \gamma} \nonumber \\
& - A_{1 \gamma} \frac{d^2 A_{1 \beta}}{dN^2} - A_{11} A_{1 \gamma} \frac{dA_{1 \beta}}{dN}
 - A_{11}A_{1 \gamma} \sum ^4 _{d=2} A_{1d}A_{d \beta} \nonumber \\
& - 2A_{1 \gamma} \sum ^4 _{d=2} \frac{dA_{1 d}}{dN}A_{d \beta} -A_{1 \gamma} \sum ^4 _{d=2} A_{1d} 
\frac{dA_{d \beta}}{dN} - A_{1 \gamma} \sum ^4 _{d,e=2} A_{1e}A_{ed}A_{d \beta} \bigg )\, , \\
M_1 \equiv &- 3 \frac{d^2 A_{11}}{dN^2} -3 \sum ^4 _{\alpha =2} \frac{dA_{1 \alpha}}{dN} A_{\alpha 1} 
 -2 \sum ^4 _{\alpha = 2} A_{1 \alpha} \frac{dA_{\alpha 1}}{dN} - \sum ^4 _{\alpha , \beta =2} A_{1 \beta}
A_{\beta \alpha}A_{\alpha 1} \nonumber \\ 
& - \sum_{\alpha , \beta , \gamma =2 \& \alpha \neq \beta \neq \gamma}^4 D(\alpha , \beta , \gamma)
\bigg \{ \bigg ( 2A_{1 \beta} \frac{dA_{11}}{dN} +A_{1 \beta} \sum ^4 _{d=2} A_{1d}A_{d1} 
+ \frac{d^2 A_{1 \beta}}{dN^2} \nonumber \\
&  +2 \sum ^4 _{d=2} \frac{dA_{1 d}}{dN} A_{d \beta} + \sum ^4 _{d=2}A_{1d} \frac{dA_{d \beta}}{dN} 
+ \sum ^4 _{d,e=2}A_{1e}A_{ed}A_{d \beta} \bigg ) 
\bigg ( A_{1 \beta} \frac{dA_{1 \gamma}}{dN} \nonumber \\
& + A_{1 \beta} \sum ^4 _{d=2}A_{1d}A_{d \gamma} 
 -A_{1 \gamma} \frac{dA_{1 \beta}}{dN} - A_{1 \gamma} \sum ^4 _{d=2}A_{1d}A_{d \beta} \bigg ) \nonumber \\
& - \bigg ( \frac{dA_{1 \beta}}{dN} + \sum ^4 _{d=1} A_{1d}A_{d \beta} \bigg )
\bigg ( A_{1 \beta} \frac{d^2 A_{1 \gamma}}{dN^2} + 2A_{1 \beta} \sum ^4 _{d=2} \frac{dA_{1d}}{dN}A_{d \gamma} \nonumber \\
& + A_{1 \beta} \sum ^4 _{d=2} A_{1d} \frac{dA_{d \gamma}}{dN} 
+ A_{1 \beta} \sum ^4 _{d,e=2} A_{1e}A_{ed}A_{d \gamma } - A_{1 \gamma} \frac{d^2 A_{1 \beta}}{dN^2} \nonumber \\
& - 2A_{1 \gamma} \sum ^4 _{d=2} \frac{dA_{1d}}{dN}A_{d \beta} - A_{1 \gamma} \sum ^4 _{d=2} A_{1d} \frac{dA_{d \beta}}{dN} 
 - A_{1 \gamma} \sum ^4 _{d,e=2} A_{1e}A_{ed}A_{d \beta }\bigg ) \bigg \}\, , \\
M_0 \equiv &- \frac{d^3 A_{11}}{dN^3} -3 \sum ^4 _{\alpha = 2} \frac{d^2 A_{1 \alpha}}{dN^2}A_{\alpha 1}
 -3 \sum ^4 _{\alpha = 2} \frac{dA_{1 \alpha}}{dN} \frac{dA_{\alpha 1}}{dN} - \sum ^4 _{\alpha = 2}A_{1 \alpha}
\frac{d^2 A_{\alpha 1}}{dN^2} \nonumber \\
& - 3 \sum ^4 _{\alpha , \beta = 2} \frac{dA_{1 \beta}}{dN} A_{\beta \alpha}A_{\alpha 1} 
 - 2 \sum ^4 _{\alpha , \beta =2} A_{1 \beta} \frac{dA_{\beta \alpha}}{dN} A_{\alpha 1}
 - \sum ^4 _{\alpha , \beta = 2} A_{1 \beta} A_{\beta \alpha} \frac{dA_{\alpha 1}}{dN} \nonumber \\
& - \sum ^4 _{\alpha , \beta , \gamma = 2}
A_{1 \gamma} A_{\gamma \beta} A_{\beta \alpha} A_{\alpha 1} 
+ \sum_{\alpha , \beta , \gamma =2 \& \alpha \neq \beta \neq \gamma}^4 D(\alpha , \beta , \gamma) \bigg \{ 
\bigg ( A_{11} \frac{d^2 A_{1 \beta}}{dN^2} \nonumber \\
& + 2A_{11} \sum_{d=2}^4 \frac{dA_{1d}}{dN}A_{d \beta}
+ A_{11} \sum_{d=2}^4 A_{1d} \frac{dA_{d \beta}}{dN} 
+ A_{11} \sum_{d,e=2}^4 A_{1e}A_{ed}A_{d \beta} \nonumber \\
& -A_{1 \beta} \frac{d^2 A_{11}}{dN^2} - 2A_{1 \beta} \sum_{d=2}^4 \frac{dA_{1d}}{dN}A_{d 1}
 - A_{1 \beta} \sum_{d=2}^4 A_{1d} \frac{dA_{d1}}{dN} \nonumber \\
&- A_{1 \beta} \sum_{d,e=2}^4 A_{1e}A_{ed}A_{d1} \bigg )
\bigg ( A_{1 \beta} \frac{dA_{1 \gamma}}{dN} + A_{1 \beta} \sum^4 _{f=2} A_{1f}A_{f \gamma}-A_{1 \gamma}
\frac{dA_{1 \beta}}{dN} \nonumber \\
& - A_{1 \gamma} \sum ^4 _{f=2} A_{1f}A_{f \beta} \bigg ) 
 - \bigg ( A_{1 \beta} \frac{d^2 A_{1 \gamma}}{dN^2} + 2A_{1 \beta} \sum_{d=2}^4 \frac{dA_{1d}}{dN}A_{d \gamma} \nonumber \\
& + A_{1 \beta} \sum_{d=2}^4 A_{1d} \frac{dA_{d \gamma}}{dN} + A_{1 \beta} \sum_{d,e=2}^4 A_{1e}A_{ed}A_{d \gamma}
 -A_{1 \gamma} \frac{d^2 A_{1 \beta}}{dN^2} \nonumber \\ 
& - 2A_{1 \gamma} \sum_{d=2}^4 \frac{dA_{1d}}{dN}A_{d \beta}
 - A_{1 \gamma} \sum_{d=2}^4 A_{1d} \frac{dA_{d \beta}}{dN} 
 - A_{1 \gamma} \sum_{d,e=2}^4 A_{1e}A_{ed}A_{d \beta} \bigg ) \nonumber \\
& \times \bigg ( A_{1 1} \frac{dA_{1 \beta}}{dN} + A_{1 1} \sum^4 _{f=2} A_{1f}A_{f \beta}-A_{1 \beta}
\frac{dA_{1 1}}{dN} - A_{1 \beta} \sum ^4 _{f=2} A_{1f}A_{f 1} \bigg ) \bigg \}\, ,\\
D(\alpha , \beta , \gamma) \equiv & \frac{1}{2}
\bigg \{ \bigg ( A_{1 \alpha} \frac{d^2 A_{1 \beta}}{dN^2} + 2A_{1 \alpha} \sum_{d=2}^4 \frac{dA_{1d}}{dN}A_{d \beta}
+ A_{1 \alpha} \sum_{d=2}^4 A_{1d} \frac{dA_{d \beta}}{dN} \nonumber \\
& + A_{1 \alpha} \sum_{d,e=2}^4 A_{1e}A_{ed}A_{d \beta} 
 - A_{1 \beta} \frac{d^2 A_{1 \alpha}}{dN^2} - 2A_{1 \beta} \sum_{d=2}^4 \frac{dA_{1d}}{dN}A_{d \alpha} \nonumber \\
& - A_{1 \beta} \sum_{d=2}^4 A_{1d} \frac{dA_{d \alpha}}{dN} - A_{1 \beta} \sum_{d,e=2}^4 A_{1e}A_{ed}A_{d \alpha} \bigg )
\bigg ( A_{1 \beta} \frac{dA_{1 \gamma}}{dN} \nonumber \\
& + A_{1 \beta} \sum^4 _{f=2} A_{1f}A_{f \gamma}-A_{1 \gamma}
\frac{dA_{1 \beta}}{dN} - A_{1 \gamma} \sum ^4 _{f=2} A_{1f}A_{f \beta} \bigg ) \nonumber \\
& - \bigg ( A_{1 \beta} \frac{d^2 A_{1 \gamma}}{dN^2} + 2A_{1 \beta} \sum_{d=2}^4 \frac{dA_{1d}}{dN}A_{d \gamma}
+ A_{1 \beta} \sum_{d=2}^4 A_{1d} \frac{dA_{d \gamma}}{dN} \nonumber \\
& + A_{1 \beta} \sum_{d,e=2}^4 A_{1e}A_{ed}A_{d \gamma}
 -A_{1 \gamma} \frac{d^2 A_{1 \beta}}{dN^2} - 2A_{1 \gamma} \sum_{d=2}^4 \frac{dA_{1d}}{dN}A_{d \beta} \nonumber \\ 
& - A_{1 \gamma} \sum_{d=2}^4 A_{1d} \frac{dA_{d \beta}}{dN} 
 - A_{1 \gamma} \sum_{d,e=2}^4 A_{1e}A_{ed}A_{d \beta} \bigg ) \bigg ( A_{1 \alpha} \frac{dA_{1 \beta}}{dN} \nonumber \\
& + A_{1 \alpha} \sum^4 _{f=2} A_{1f}A_{f \beta}-A_{1 \beta}
\frac{dA_{1 \alpha}}{dN} - A_{1 \beta} \sum ^4 _{f=2} A_{1f}A_{f \alpha} \bigg ) \bigg \} ^{-1}
\nonumber \\
& \times \bigg ( \frac{d^3 A_{1 \alpha}}{dN^3} +3 \sum ^4 _{d=2} \frac{d^2 A_{1d}}{dN^2} A_{d \alpha}
+ 3 \sum ^4 _{d=2} \frac{dA_{1d}}{dN} \frac{dA_{d \alpha}}{dN} +\sum ^4 _{d=2} A_{1d} \frac{d^2 A_{d \alpha}}{dN^2}
\nonumber \\
& + 3 \sum ^4 _{d,e=2} \frac{dA_{1e}}{dN}A_{ed}A_{d \alpha}+2 \sum ^4 _{d,e=2} A_{1e} \frac{dA_{ed}}{dN}A_{d \alpha}
+ \sum ^4 _{d,e=2} A_{1e}A_{ed} \frac{dA_{d \alpha}}{dN} \nonumber \\
& + \sum ^4 _{d,e,f=2} A_{1f}A_{fe}A_{ed}A_{d \alpha} \bigg )\, .
\label{1450}
\end{align}


\end{document}